\documentclass[aps,prl,twocolumn,groupedaddress]{revtex4}

\usepackage{graphicx}% Include figure files
\usepackage[colorlinks]{hyperref}
\hypersetup{urlcolor=blue,linkcolor=blue,citecolor=blue,filecolor=blue} % to use with revtex4
\usepackage{color,soul}
\sethlcolor{yellow}

\begin{document}

\title{Taming the resistive switching in Fe/MgO/V/Fe magnetic tunnel
junctions:\\ An ab initio study}
%\shorttitle{Resistive switching in Fe/MgO/V/Fe}
 
\author{J.M. Aguiar-Hualde}
%\email{aguiar@ipcms.u-strasbg.fr}
%\homepage{http://legauss.blogspot.com}
% \address{IPhT, CEA/Saclay, Orme des Merisiers, 91190 Gif-sur-Yvette Cedex,
% France.}
\affiliation{IPhT, CEA/Saclay, Orme des Merisiers, 91190 Gif-sur-Yvette Cedex,
France.}
\author{M. Alouani}
%\email{mea@ipcms.u-strasbg.fr}
% \address{IPCMS, UMR 7504 CNRS-UdS, 23 rue du Loess, Strasbourg 67034, France.}
\affiliation{IPCMS, UMR 7504 CNRS-UdS, 23 rue du Loess, Strasbourg 67034, France.}
 
%\date{\today}
%{Density functional theory, local density approximation, gradient correction}
%{Total energy calculations}
%{Spin transport through interfaces}
 
\begin{abstract}
A possible mechanism for the resistive switching observed experimentally in
Fe/MgO/V/Fe junctions is presented. Ab initio total energy calculations within
the local density approximation and pseudopotential theory shows that by moving
the oxygen ions across the MgO/V interface one obtains a metastable state. It
is argued that this state can be reached by applying an electric field across
the interface. In addition, the ground state and the metastable state show
different electric conductances. The latter results are discussed in terms of
the changes of the density of states at the Fermi level and the charge transfer
at the interface due to the oxygen ion motion.
\end{abstract}
 
%\pacs{71.15.Mb, 71.15.Nc, 72.25.Mk} \keywords{switching}
\maketitle

\section{Introduction} 

Bipolar nonvolatile Resistive Switching (RSw), is one of the most largely
studied phenomena \cite{Sawa2008, Waser2007,Szot1992,Yang2008} for the
development of next generation of random access memory (RAM) devices. Some of
these devices have a metal-insulator-metal (M/I/M) structure like graphene
oxides (GO) of the form Cu/GO/Pt. In these systems, the RSw was explained
\cite{He2009} in terms of desorption/absorption of oxygen-related groups on the
GO sheets and the diffusion on the top Cu electrodes. MgO-based magnetic-tunnel
junctions (MTJ) are also good candidates for RAM memories since it was shown
that they have good RSw reproducibility \cite{Halley2008}. One of the aspects
that makes them particular, is that they can exhibit two kinds
\cite{Teixeira2009} of switching: magnetoresistive and structural (RSw). This
enlarge the scope of applicability of these memories.

Despite the great interest on the tunneling magnetoresistance (TMR) features of
these kind of junctions due to the large values predicted by Butler and
coworkers for Fe/MgO/Fe \cite{Butler2001}, the mechanism for the RSw phenomenon
remains to be elucidated. However, in the last few years, several possible
mechanisms have been proposed \cite{Waser2009}. For example, to explain the
bipolar RSw, three different effects are mentioned in the literature. The first
one is the electrostatic/electronic effects, the second the electrochemical
metalization (ECM) effect, and the third the valence change memory (VCM)
effect. The first effect is based on purely electronic phenomena, like those
induced by charge traps \cite{Simmons1967, Rozenberg2004}.  The other two
effects involve oxidation-reduction (redox) processes. In ECM, ions travel
across the insulating spacer from the active electrode to the inert electrode,
developing dendrites, while in VCM, the migration of anions takes place when
applying a voltage pulse and produces reduction or oxidation reactions
depending on the polarity \cite{Szot1992,Yang2008}. According to the theory of
filamentary conduction \cite{Dearnaley1970}, the redox reactions take place
across the filaments grown along defects in the insulating spacer.  The size of
the MgO spacer seems also to play an important role \cite{Teixeira2009}, e.g.,
thin MTJs usually exhibit RSw from the virgin state, while thick ones go
through an electroforming process (in which filaments are developed) before
displaying the phenomenon. Some models \cite{Rozenberg2010} have been proposed
to simulate the filamentary conduction on transition-metal oxides.

The models that describe the RSw, rely usually on an assumption of two local
minima in the energy profile. Recently, a double well model for trapped
electrons in MgO-based tunnel junctions \cite{Bertin2011} confirm the power law
dependencies of resistance observed experimentally \cite{Najjari2010}. 

Using first principle transport calculation\cite{Bose2008,Heiliger2007}, it is
shown that the tunneling magnetoresistance ``TMR'' of Fe/MgO/Fe junctions is
strongly diminished by small oxygen concentrations in a single partially
oxidized FeO interface layer. This reduction is attributed to the reduction of
specular contributions to the conductances of the parallel configuration of the
lead magnetizations highlighting the importance of ordered interfaces for large
TMR ratios. This later conclusion is also in agreement with that of Wortamnn
\textit{et al.}\cite{Wortmann2004} regarding the interplay between electronic
structure, atomic structure and the tunneling process. The barrier thickness in
the limit of coherent tunneling is also shown to be
important\cite{Heiliger2008} since it is shown that at large barrier
thicknesses, only a small amount of states contributes to the overall current.
Another important issue concerns the interchannel diffusive scattering by
disordered oxygen vacancies located at or near the Fe/MgO interface which is
shown to drastically reduce the tunnel the TMR from the ideal theoretical limit
to the presently observed much smaller experimental range.\cite{Ke2010} 

In the Fe/MgO/Fe junctions, the effect of the inclusion of different transition
metal atomic species at one of the Fe/MgO interface has been studied. For
example, chromium or vanadium were inserted as an electron symmetry filter in
the system, \cite{Halley2008,Najjari2010}. Both chromium and vanadium can be
easily polarized by proximity with a magnetic material such as iron. The
memristor model \cite{Chua1971} was shown to be good for explaining the
hysteresis-like I-V characteristics. The conductance of Fe/MgO/V/Fe was shown
to oscillate with increasing the number of vanadium layers \cite{Feng2009}.
Another reason for including vanadium at the Fe/MgO interface, lies in the
higher oxygen affinity of this element compared to iron.  This would make it
possible for the appearance of oxygen vacancies at the interface and would
break the symmetry of the sample defining the active electrode. This is one of
the aspects which lead to the bipolar behavior of the switching.

Usually, V is assumed to lie on top of O when grown on MgO substrate, however
this was shown not to be always the case by Ikuhara and coworkers
\cite{Ikuhara1997}. In a study of the atomic and electronic structure of V/MgO
interface, they showed that the best matching between experiment and simulation
corresponds to the case when V is located on top of Mg. This geometry allows
the oxygen ions to move in straight line across the interface when redox
reactions take place.

In this work, we focus on the study of the RSw on Fe/MgO/V/Fe with the goal of
gaining insight into the mechanism of this interesting phenomenon. To this end,
we performed our study by computing the total energy of the system for various
motions of the oxygen ions at the MgO/V interface and restrict ourselves to the
case of ferromagnetic alignment of the iron electrodes. The main goal pursued
in this work is to show whether it is possible to find two local minima in the
total energy profile by identifying two possible positions for an oxygen ion
moving across the MgO/V interface. This will simulate the oxidation which is
assumed to take place in the RSw phenomenon as suggested by experiment in the
case of KNbO$_3 $\cite{Szot1992}. 

The geometrical configuration of the system with the oxygen ion in the MgO
layer, is referred to as the initial state configuration, while the final
configuration corresponds to the oxygen in the position identified by the
minimum of the total energy. To achieve this goal, we tried several types of
atomic arrangements in order to contrast the results and optimize the way to
simulate the oxidation. The zero bias electrical conductance is then computed
and the results are discussed in reference to those obtained experimentally.
Notice however that the experimental thickness of the MgO insulating barrier of
about 3 nm is much larger that the one used in our calculation which is less
than one nm. Due to this sever limitation, our calculations can only be
compared qualitatively to the experimental results.

\section{Computational Details} 

The electronic conductance was calculated using the SMEAGOL \cite{Rocha2006}
package where the non-equilibrium Green's function formalism is employed. In
this code, the ballistic regime is assumed so the Landauer-B\"{u}ttiker formula
for the current is valid for computing the electric current. To compute the
conductance, we have used 400$\times$400 {\bf k} points in the two dimensional
Brillouin zone as used in Ref. \cite{Feng2009}.  This high mesh is required for
the good convergence of the conductance. The electronic structure required for
the calculations, is performed by means of the SIESTA code \cite{Soler2002}
which is a density functional theory (DFT) code based on the pseudopotential
approximation and localized numerical orbital basis set. Apart from specifying
the structural parameters of our system, we have to chose the type of
exchange-correlation (XC) functional, the pseudopotentials (PPs) for the atomic
species, and the size of the basis set.  For this study, the PPs were generated
using the improved Troullier-Martins scheme\cite{Troullier1991}. The local
density approximation (LDA) was employed for both the generation of the
pseudopotentials and the exchange-correlation potential \cite{Perdew1981}.
Non-linear exchange-correlation core corrections were used for iron, vanadium
and magnesium and no core corrections for oxygen. The electronic population and
the cutoff radii for each atomic species and angular momentum are listed in
Table~{\ref{tab:PPs}}.

{%
\begin{table}[ht]
\begin{tabular}{ c|cccc }
\hline
\hline
 Orbital & Fe & V & Mg & O \\ \hline
$s$ &\rule{1ex}{0pt} $4s^2$ 2.00 \rule{1ex}{0pt}&\rule{1ex}{0pt} $4s^2$ 2.97 \rule{1ex}{0pt}&\rule{1ex}{0pt} $3s^2$ 2.50 \rule{1ex}{0pt}&\rule{1ex}{0pt} $2s^2$ 1.25 \rule{1ex}{0pt} \\
$p$ & $4p^0$ 2.24 & $4p^0$ 3.50 & $3p^0$ 2.50 & $2p^4$ 1.25 \\
$d$ & $3d^6$ 1.78 & $3d^3$ 1.59 & $3d^0$ 2.50 & $3d^0$ 1.25 \\
$f$ & $4f^0$ 2.00 & $4f^0$ 2.17 & $4f^0$ 2.50 & $4f^0$ 1.25 \\
\hline
\hline
\end{tabular}
\caption {Electronic configuration (left) and cutoff radii for each atomic
species and angular momentum.}
\label{tab:PPs}
\end{table}
}%

For the basis set, a split-valence type was used with a split norm value of
0.15. Double-$\zeta$ type was employed for all the atoms and angular momentum
while polarization functions was used only for iron, vanadium and magnesium.
The values chosen for the cutoff radii of the first $\zeta$ are summarized in
Table~{\ref{tab:Basis}}. The values for the cutoff radii of the second $\zeta$
were left to be determined by the split norm value. The electronic temperature
and the mesh cutoff are set to 300K and 350 Ry, respectively. The structural
parameters were those of Butler and coworkers \cite{Butler2001}: iron sits on
top of oxygen with a distance of 2.16 \AA{}, vanadium occupies the place of
iron in the interface V/MgO. The iron lattice constant is 2.866 \AA{} and MgO
lattice constant is $\sqrt{2}$ times that of the Fe. As mentioned before,
following the work of Ikuhara and coworkers on V/MgO interfaces
\cite{Ikuhara1997} we have also studied the positioning of V on top of Mg.

{%
\begin{table}[ht]
\begin{tabular}{ c c c c }
\hline
\hline
Fe & V & Mg & O \\ \hline
4$s$: $r_c = 6.0$ \rule{1ex}{0pt}& 4$s$: $r_c = 6.0$ \rule{1ex}{0pt}& 3$s$: $r_c = 6.0$ \rule{1ex}{0pt}& 2$s$: $r_c = 4.0$ \\ 
3$d$: $r_c = 5.0$ \rule{1ex}{0pt}& 3$d$: $r_c = 5.0$ \rule{1ex}{0pt}& \rule{1ex}{0pt}& 2$p$: $r_c = 4.5$ \\
\hline
\hline
\end{tabular}
\caption {Cutoff radii for the first $\zeta$ basis function for each atom and
angular momentum.} 
\label{tab:Basis}
\end{table}
}%

\begin{figure}[ht!]
\begin{center}
\includegraphics[width=\columnwidth]{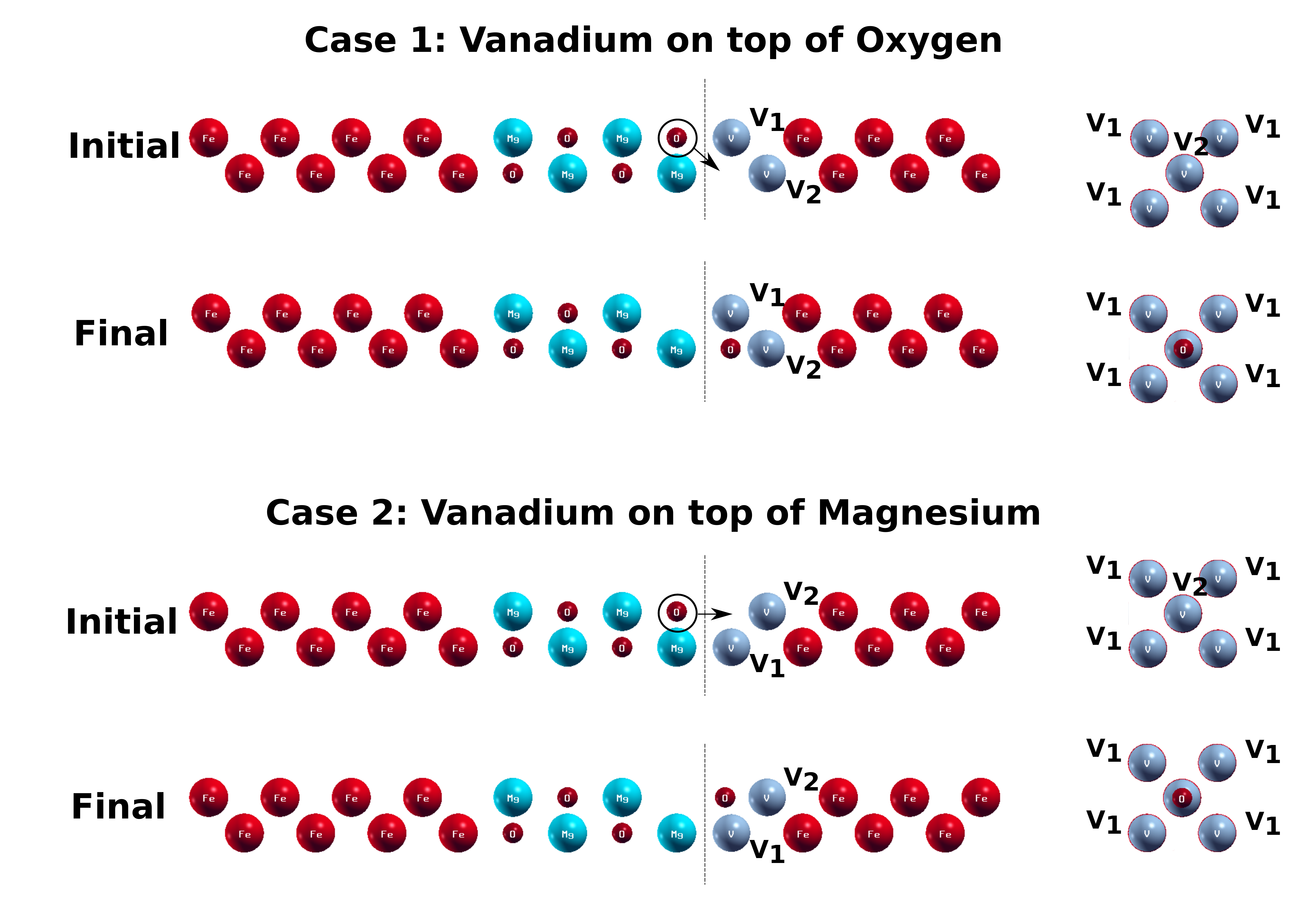}
\end{center}
\caption {Initial and final unit cells for the two cases considered: the
vanadium is on top of oxygen (top drawings) and V on top of Mg (bottom
drawings). The motion of oxygen ion is shown by an arrow. The first and second
layer of vanadium (V$_1$ and V$_2$) are identified for further reference. For
clarity, the interface  at the vanadium 1 is shown on the right in $xy$ plane
for the initial and final states.}
\label{fig:Structures} 
\end{figure}

%\section{Results}

\section{Results and discussion} 

\subsection{Total energy local minimum due to oxygen motion} 

The motion of oxygen ions may take place when defects or oxygen vacancies are
present \cite{Szot1992}. This could be the reason for the none existence of
resistive switching in Fe/MgO/Fe systems: introducing vanadium in one of the
Fe/MgO interfaces, defects and dislocations are more likely to appear, and in
this way, negatively charged oxygen ions have more space where to migrate under
an applied electric field.

\begin{figure}[ht!]
\begin{center}
\includegraphics[width=\columnwidth]{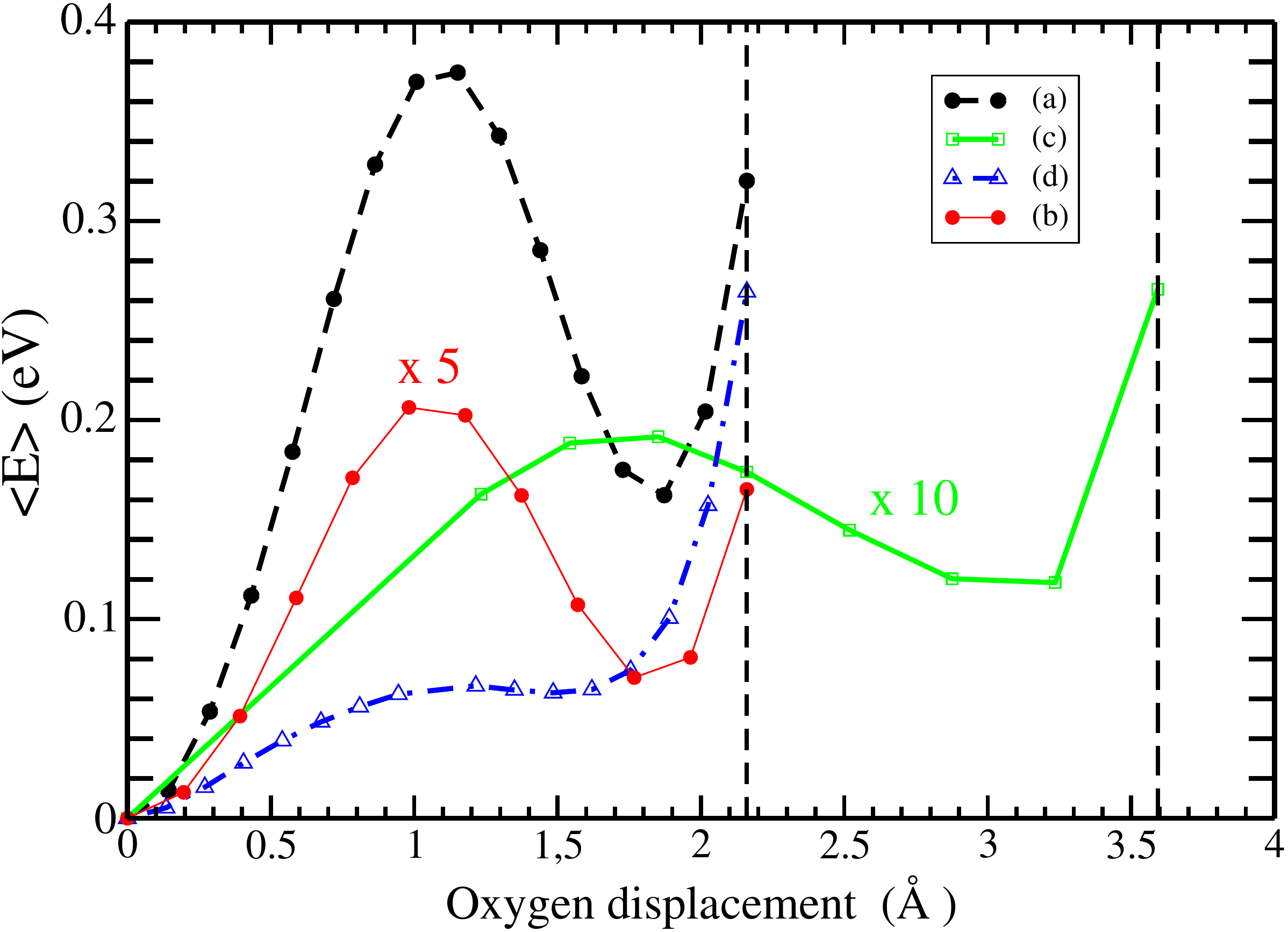}
\end{center}
\caption {Average total Energy per atom in eV as a function of oxygen
displacement in (\AA) for the several cases of oxygen motion : (a) small unit
cell; (b) large unit cell and only one oxygen is allowed to move; (c) large
unit cell, V vacancy; (d) small unit cell, V on top of Mg. Cases (b) and (c)
are amplified 5 and 10 times respectively. All the cases are performed within
the LDA approximation. The origin of the displacement corresponds to the
position of the last MgO layer, the dashed line at 2.16 \AA{} refers to the
first vanadium layer and the dashed line at 3.593 \AA{} to the second one.}
\label{fig:Etot}
\end{figure}
\begin{figure}[ht!]
\begin{center}
\includegraphics[width=\columnwidth]{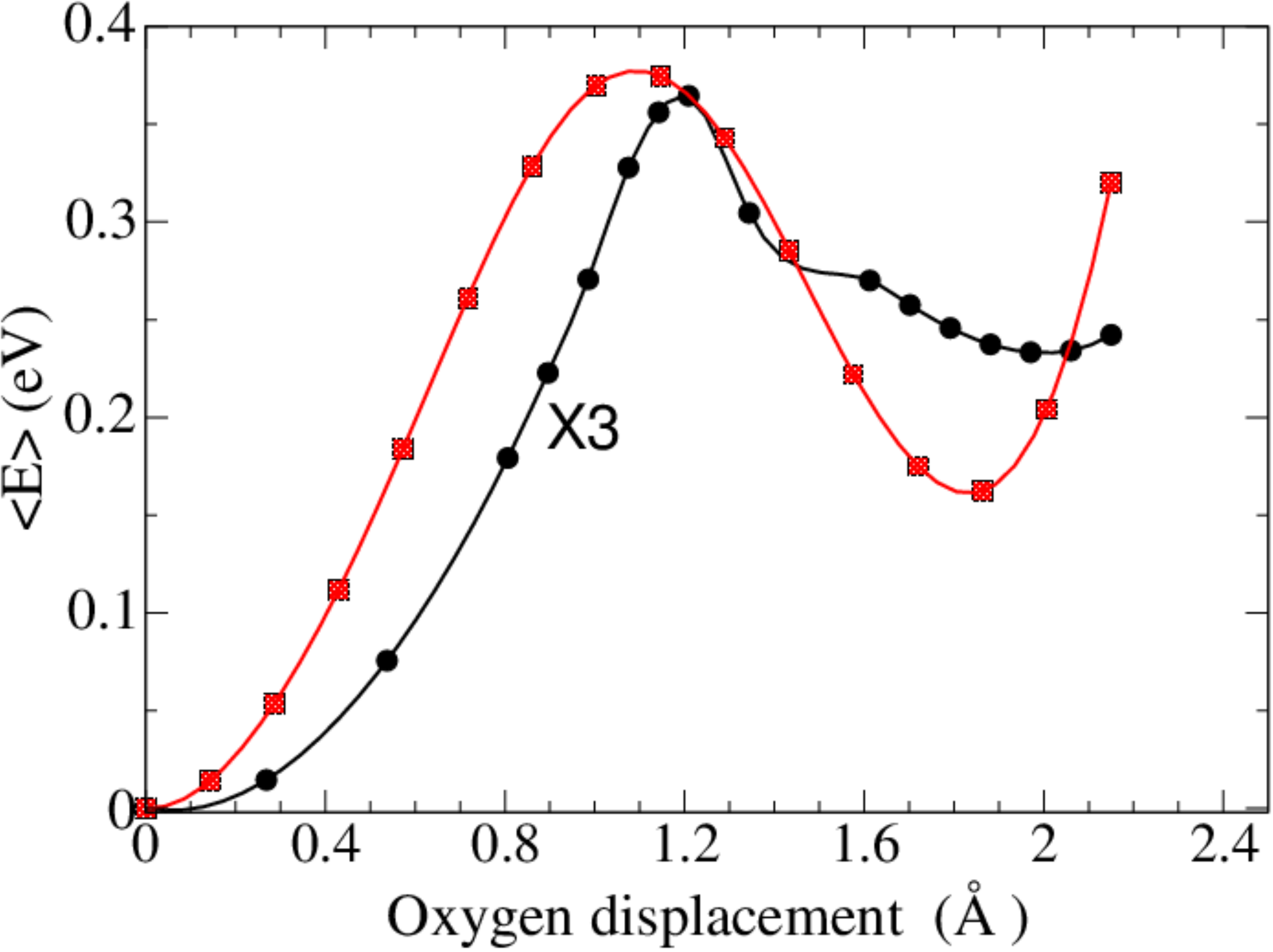}
\end{center}
\caption {Calculated average total Energy per atom in eV as a function of
oxygen displacement in (\AA) with total atomic relaxation (black dots) compared
to the unrelaxed structure (red dots). Notice that the atomic relaxation
reduces drastically the energy barrier height.}
\label{fig:relax}
\end{figure}
% 
% Directories for the figures:
% (a)./Twelve/07-LDA/16.Points/??/FERRO ./Doc_smalley/03-Strasbg/20-LDA.Small/
% (b)./Twelve/12-FVMF.3x3/??/
% (c)./Twelve/11-V.Vacancy.Bulk/*/
% (d)./Eniac/08-Vanadium.Vacancy.3x3/FVMF??/FERRO/
% (e)./Doc_smalley/03-Strasbg/06-FeVMgO/06-DiezCorridas/Eniac/RSLT??/
% (f)./Twelve/10-New.Geometry/??/PP/
% 
% Data:
% ./Doc_smalley/03-Strasbg/06-FeVMgO/10-Results.Summarized
% ./Twelve/00-ANALYSIS/

Fig. \ref{fig:Structures} depicts the initials and final configurations on the
2 cases considered in this work: vanadium on top of oxygen and vanadium on top
of magnesium. Fig. \ref{fig:Etot} shows the results for the average total
energy per atom as a function of the oxygen position (only the $\hat{z}$
component, orthogonal to the interface planes) for all the cases considered. In
order to compare the results, all the curves are relatives to the energy of the
initial configuration, i.e. when the oxygen ion is in its initial state. Curve
(a) corresponds to the case taken as a reference for the other possible paths.
In this case, we considered a relatively small unit cell (24 atoms) and the
oxygen ion in front of the vanadium layer is moved to oxidize that layer. It is
worth noticing that the motion of the oxygen ion, is not along $\hat{z}$
direction but diagonal when there is a vanadium atom in front as is shown in
figure \ref{fig:Structures}. Here we observed one of the main results of this
work: somewhere near 1.85 \AA{} from the initial position, the total energy
exhibits a minimum. It can be argued that the barrier energy is very high of
about 0.37 eV but this is a consequence of this kind of motion, where one
oxygen ion is moved in this small unit cell. This will correspond to moving the
whole oxygen layer in the last MgO layer to simulate the oxidation of vanadium.
In order to validate this hypothesis, we calculated the same kind of motion but
in a larger unit cell (215 atoms in total) by moving one out of 9 oxygen ions.
The result presented by curve (b) shows that the barrier's height diminishes by
an order of magnitude to 41 meV or when converted to temperature is about 476
K.  This is a realistic barrier that can be overcome easily by an electric
field. We will now explore other possible oxygen motions to reduce further the
barrier height. 

% We have explored different kind of variations in order to contrast results:
% vanadium vacancy in (c) and (d), GGA exchange-correlation functional in (e) and
% a different geometry (V on top of Mg instead of oxygen) in (f). This last case
% corresponds to the geometry which Ikuhara et coworkers \cite{Ikuhara1997}
% consider more likely to have for MgO/V interface. Furthermore, it is worth
% noticing that the energy barrier height is the lowest, closer to realistic values.
% Vanadium vacancies make the minimum to shift deeper into the vanadium layers
% while considering vanadium on top of magnesium (as suggested by Ikuhara and coworkers
% \cite{Ikuhara1997}) lowers the potential barrier.

In Fig. \ref{fig:Etot} (c), we have considered the same large unit cell as in
(b) but a vanadium vacancy in front of the oxygen. Finally, in Fig.
\ref{fig:Etot} (d) we have used the same small unit cell as in (a) but V on top
of Mg (case 2 in Fig. \ref{fig:Structures}). In these two cases, the oxygen
moves in a straight line and the last case corresponds to the geometry which
Ikuhara et coworkers \cite{Ikuhara1997} consider more likely for a MgO/V
interface. It is worth noticing that in both cases the energy barriers height
diminish drastically being (c) the lowest, closer to realistic values (0.017 eV
or about 197 K). Vanadium vacancies make the minimum to shift deeper into the
vanadium layers while considering V on top of Mg (as suggested by Ikuhara and
coworkers \cite{Ikuhara1997}) lowers the potential barrier. However, the second
minimum at 1.5\AA, is very shallow and is about 4 meV lower that the barrier
height of 66.5 meV (see \ref{fig:Etot}d). This shallow minimum of about 46 K
can trap an oxygen ion only at low temperatures.  The comparison of all these
total energy calculations as a function of the position of one of the oxygen
ions suggests that case (a) and (d) are the most representative of oxygen ion
migration across the interface. Despite the different ways that the oxygen ion
might migrate across the interface or how its motion is computed, a general
feature emerges. There is a local energy minimum when an oxygen ion is moved
across the interface. We argue that because the oxygen ion is electrically
charged, such a motion can be triggered by an applied strong electric field
pulse. The trapped negatively charge oxygen ion in this minimum can reverse
motion when the electric field is applied in the opposite direction. 

There are two major factors that might reduce drastically the calculated energy
barrier heights. The first one is the atomic relaxations around the displaced
oxygen ion which are not taken into account due to high computational cost.
Nevertheless, we have performed atomic relaxation for the small unit cell in
absence of an applied electric field, which is not so computationally difficult
to implement. We compare in Fig. \ref{fig:relax} the average total energy per
atom as a function of the oxygen displacement for the relaxed and the not
relaxed calculations. We can see that the atomic relaxations reduce drastically
the energy barrier by enhancing the electronic screening around the displaced
oxygen ion. However, we believe that such huge relaxation are overestimated
because in presence of an applied electric field, the atoms around the
displaced oxygen atoms will have less time to relax. We expect therefore that
the atomic relaxations will not play a very important role. The Second factor
that might reduce the barrier height involves oxygen vacancies. Experiment
shows that positively charged oxygen vacancies are also important for the
oxygen ions diffusion under an applied electric field \cite{Szot1992,Yang2008}.
One therefore expects that an oxygen vacancy distribution would significantly
reduce the energy barrier since the negatively charged oxygen ions will have
enough space to move. Such calculations are computationally expensive and are
therefore beyond the scope of the present study.  Since the non optimized
oxygen motion produces large energy barrier heights, it is difficult to
estimate a realistic electric field that will move the oxygen ions (O2-). If we
consider that $\Delta V = d E$ (where $\Delta V$ is the potential corresponding
to the energy barrier, $E$ the applied electric field, and $d$ the displacement
of oxygen ion) then for $\Delta V$ of 1 volts and $d$ of the order of 1 \AA~,
we need a field of at least 10$^{8}$ V/m. This is a large electric field and
defect pathways and other relaxation effects are therefore necessary for
reducing the energy barrier height. Notice however that experimentally the
applied voltage is about 1 volt and since the potential drop will occur only
through the insulating region of the order of 1 nm (see Ref.
\cite{Bertin2011}), the produced electric field in MgO is of the order of
10$^9$ V/m which of about the same order of magnitude than the estimated
electric field. 

\subsection{Transmission and Conductance}

\begin{figure}[ht!]
\begin{center}
\includegraphics[width=\columnwidth]{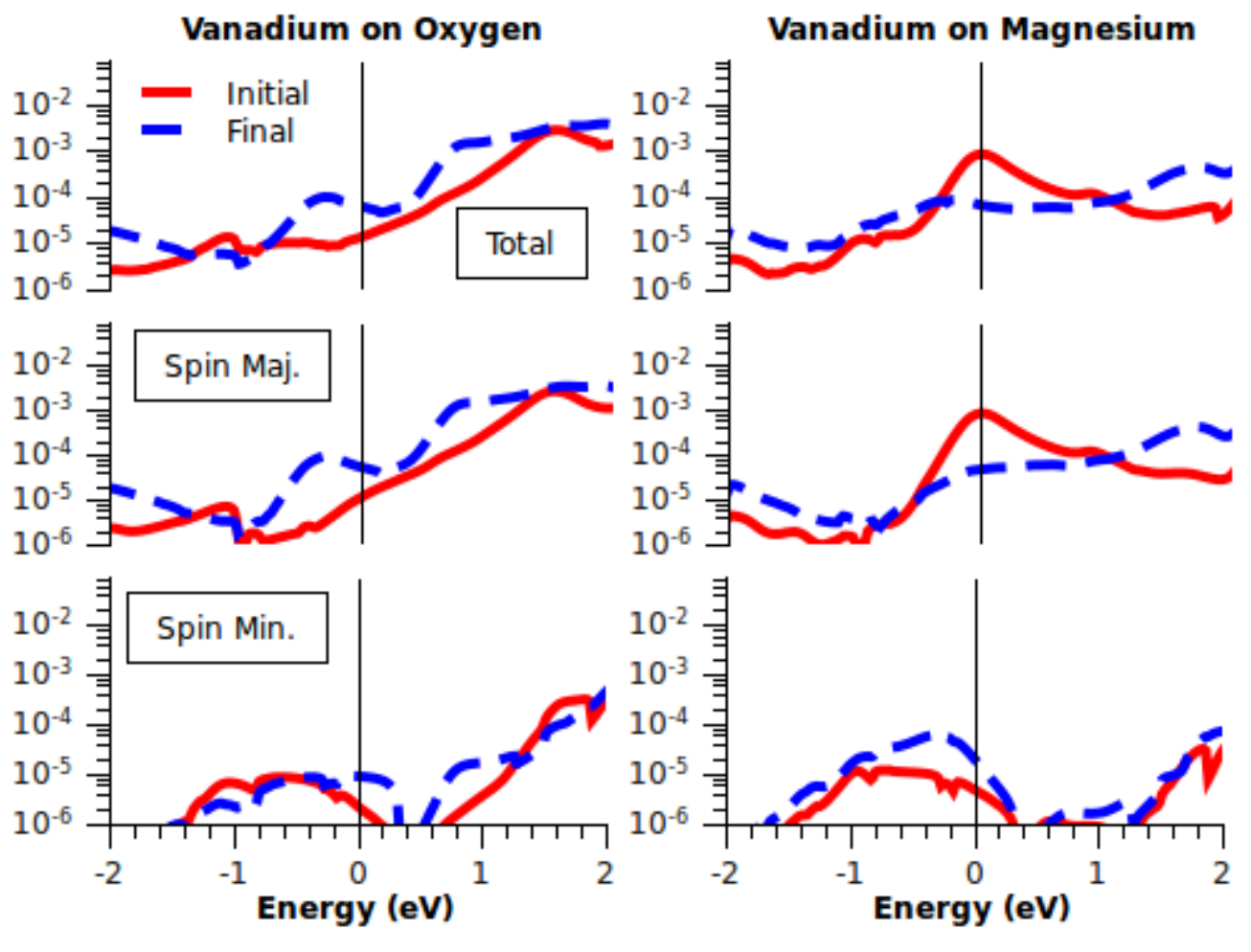}
\end{center}
\caption {Transmission as a function of energy for the initial (solid red
curves) and final (dashed green curves) configurations. Left column: V on top
of O; right column: V on top of Mg. Top row shows total transmissions; central
row the spin majority components; and the bottom row the spin minority
components. In all cases, the conductance shows a switching.}
\label{fig:TRC}
\end{figure}

In Fig. \ref{fig:TRC} we present the results for the transmission as a function
of energy at the initial and final configurations of the oxygen motion in both
studied junctions, i.e., V on top of O (left column) and V on top of Mg (right
column). Top row corresponds to the total transmission while the spin
decomposition is shown in the other two columns: spin majority ($G_{\uparrow}$)
in the center and spin minority ($G_{\downarrow}$) in the bottom. Comparing the
rows, we notice that the total transmission has almost the same shape of the
$G_{\uparrow}$ component. This is a consequence that in ours calculations we
are considering the ferromagnetic alignment for studying the RSw of the
junction. If we observe the total conductance (transmission at the Fermi level)
we find that it increases when the system goes from the initial state to the
final one in the case of V on top of O, but it has the opposite behavior in the
case of vanadium on top of magnesium. {This resistance change would be a good
test for distinguishing experimentally between the two geometries, i.e.,
correlating the sense of the switching and the orientation of the sample with
respect to the applied electric field.}

The total conductance and their spin dependent values, are summarized in table
\ref{tab:Conduc}. The labels correspond to each case in Fig. \ref{fig:Etot}.
Apart from case (c), where the total conductance has the same order of
magnitude in both initial and final states, the total conductance of the system
is one order of magnitude lower in the initial configuration than in the final
one when V is on top of O, while for V on top of Mg the values are reversed. We
notice also that the main spin component of the conductance is reversed when
comparing vanadium on oxygen and vanadium on magnesium.

% \begin{figure}[ht!]
% \begin{center}
% \includegraphics[width=\columnwidth]{fig5.pdf}
% %\includegraphics[width=0.75\textwidth]{fig4.png}
% \end{center}
% \caption {Conductance values in the initial and final configuration for all the
% LDA calculations in the ferromagnetic alignment.}
% \label{tab:Conduc}
% \end{figure}

{%
\begin{table}[ht]
\begin{tabular}{ c|cccc }
\hline
\hline
 Case & $G_{\uparrow}/G_0$ & $G_{\downarrow}/G_0$ & $G_{\tt Total}/G_0$ \\ \hline
(a) Initial &\rule{1ex}{0pt} 5.83e-05 \rule{1ex}{0pt}&\rule{1ex}{0pt} 3.02e-04 \rule{1ex}{0pt}&\rule{1ex}{0pt} 3.60e-04 \rule{1ex}{0pt} \\
(a) Final & 9.48e-04 & 1.98e-04 & 1.15e-03 \\
\hline
(b) Initial & 1.18e-03 & 4.84e-06 & 1.19e-03 \\
(b) Final & 3.51e-02 & 7.98e-06 & 3.51e-02 \\
\hline
(c) Initial & 4.59e-03 & 3.28e-06 & 4.59e-03 \\
(c) Final & 2.84e-03 & 5.13e-06 & 2.85e-03 \\
\hline
(d) Initial & 1.45e-02 & 4.52e-05 & 1.45e-02 \\
(d) Final & 5.90e-04 & 1.24e-03 & 1.83e-03 \\
\hline
\hline
\end{tabular}
\caption {Conductance values in the initial and final configurations for all
the LDA calculations in the ferromagnetic alignment. Labels correspond to those
of Fig. \ref{fig:Etot}}
\label{tab:Conduc}
\end{table}
}%

%%%%%% Analyze DOS in order to be able to say something else about the R-SW Mechanism.

\begin{figure}[ht!]
\begin{center}
\includegraphics[width=\columnwidth]{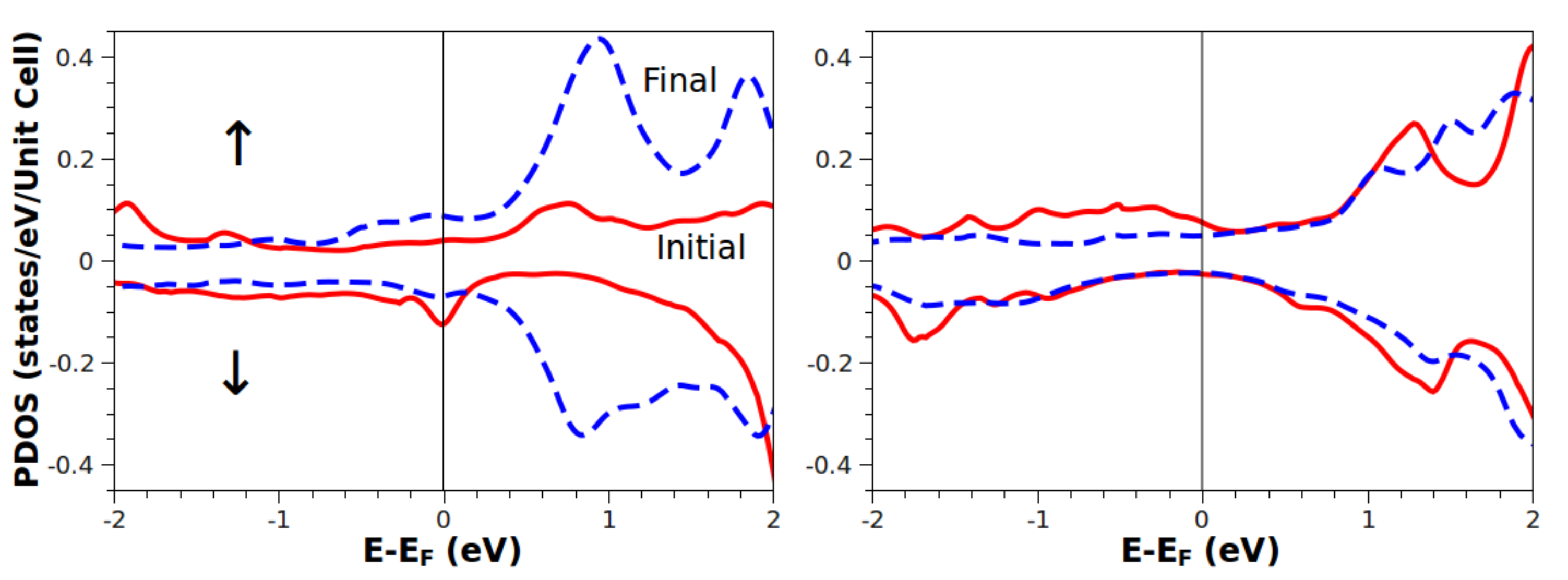}
\end{center}
\caption {$d_{3z^2 -r^2}$ component of the PDOS for the vanadium atom in front
of the moving oxygen (figure \ref{fig:Structures}): V$_1$ for V on top of O
(left panel) and V$_2$ for V on top of Mg (right panel). The Fermi level is at
the zero of energy.}
\label{fig:PDOS}
\end{figure}

\subsection{Density of states} 

From the analysis of the local density of states (PDOS) of the MgO, the highest
occupied states are at about -4 eV below the Fermi level and the lowest
unoccupied states are at about 2 eV above it. That gives an MgO band gap of
about 6 eV. This value is much smaller than the bulk value of 7.8 eV but it
seems that thin films have much smaller band gaps \cite{Kurth2006}. When the
RSw takes place, the changes in the partial density of states (PDOS) occurs
mainly for atoms located at the vicinity of the displaced oxygen ion. The most
appreciable change, corresponds to the $d_{3z^2 -r^2}$ orbital of the vanadium
in front of the oxygen in the initial position: referred to as V$_1$ for V on
top of O and V$_2$ for V on top of Mg (see figure \ref{fig:Structures}). The
results for such PDOS are shown in Fig. \ref{fig:PDOS} and explain the behavior
of the conductance discussed previously. The spin majority component of the
PDOS around the Fermi energy (taken as the energy zero), is relevant here due
to the ferromagnetic alignment of the leads. It increases in V on top of O when
the oxygen moves from the initial to final position but decreases in the same
situation for V on top of Mg. We can interpret this result in terms of the
magnetic moment and charge transfer at the MgO/V/Fe interfaces as we will see.

\begin{figure}
\begin{center}
\includegraphics[width=\columnwidth]{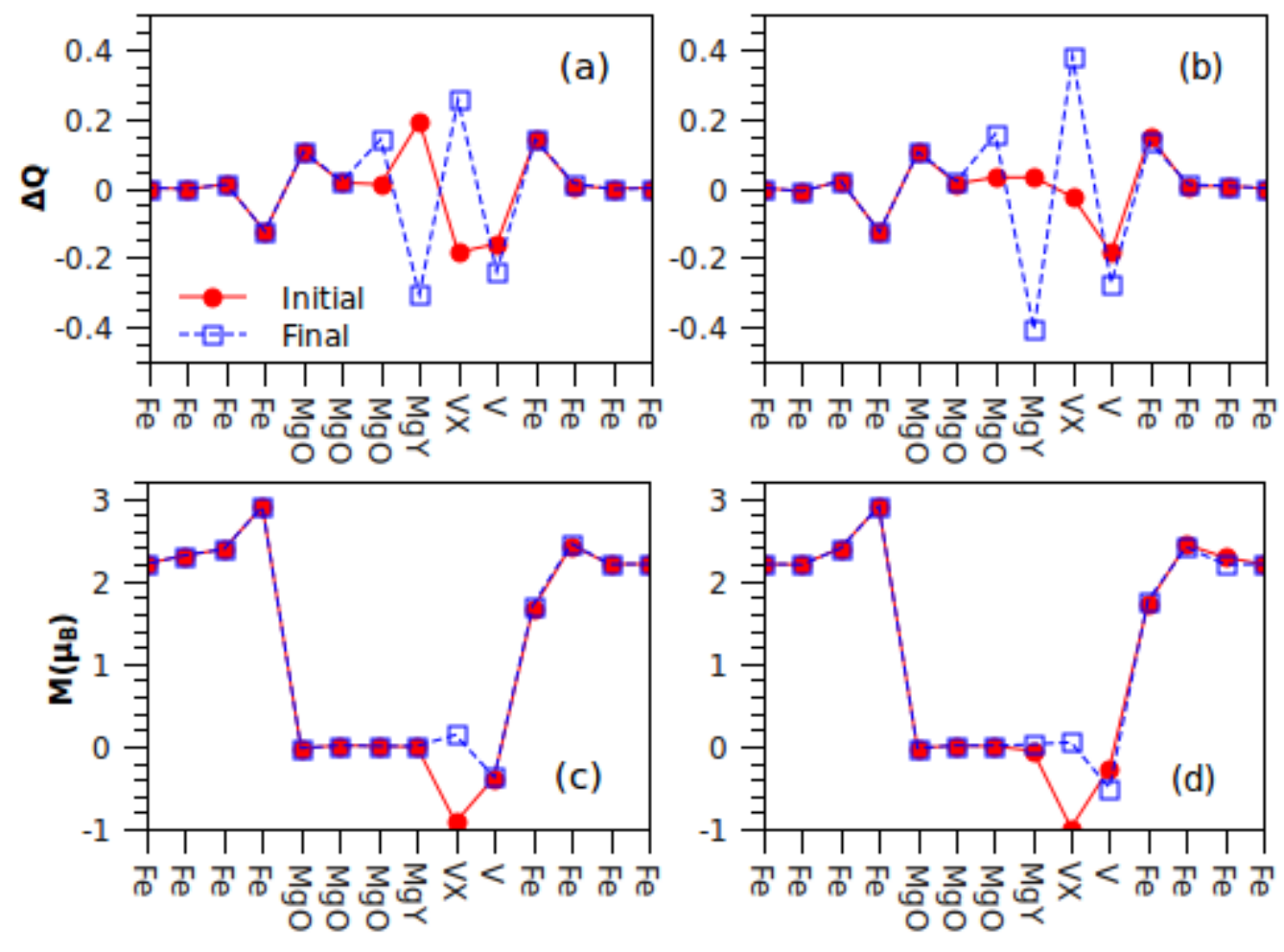}
\end{center}
\caption {Magnetic moment $\bf M$ and charge transfer $\Delta Q$ across the
system; The left (a) and (c) panels are for V on top of O and the right (b) and
(d) panels are for V on top of Mg. A positive $\Delta Q$ corresponds to a
charge accumulation on the layer and negative one to a charge depletion. In the
initial configuration X is an empty site and Y is an oxygen ion, while in the
final configurations it is the other way around.}
\label{fig:M.and.Q}
\end{figure}

\subsection{Magnetic moment and charge transfer} 

Fig. \ref{fig:M.and.Q} shows the charge transfer and the spin magnetic moments
at the Fe/MgO/V/Fe interfaces. It is clear that at the Fe/MgO interface there
is a net charge transfer from the last Fe layer towards the first MgO layer
(panels (a) and (b)), and the interface iron magnetic moment is about 3 $\mu_B$
much larger than the bulk moment of 2.2 $\mu_B$ (panels (c) and (d)).  This
moment is typical of an iron surface and is due to a lower iron coordination.
However, the MgO/V/Fe interface is much more interesting. First, when the V
atom is on top of O, it is much oxidized, i.e. there is a net charge transfer
from the vanadium layers towards both MgO and Fe (initial case in panel (a)).
But, when one oxygen ion passes into the first V plan (final case in panel
(a)), the first V layer adjacent to Mg gains about 0.25 electrons and the Mg0
layer loses about -0.3 electrons. This is essentially due to the extra charge
of the displaced oxygen ion. When V is on top of Mg (panel (b)), the situation
is a bit more complex in the final state. Due to the extra charge of the oxygen
ion, the interface vanadium layer gains about 0.4 electron while the adjacent
MgO layer loses about the same amount of charge and the second MgO layer gains
about 0.15 electron. The oxygen in the vanadium layer plays the role of an
impurity and creates a Friedel charge oscillation around it. 

The opposite behavior observed in the conductance for both configurations in
Table \ref{tab:Conduc}, can be understood by analyzing the charge transfer
presented in Fig. \ref{fig:M.and.Q}: for V on Mg (panel (a)) we have less
charge oscillation in the initial configuration than the final configuration.
This qualitatively favors a higher conductance for the initial configuration as
shown in Fig. \ref{fig:TRC}. For V on top of O, the situation is different
since the charge oscillation is about the same for the initial and final
configuration. However, we notice that both in the initial configuration both
vanadium layers have a charge depletion which amount to higher potential
barrier than in the final configuration. This might qualitatively favor a
higher conductance for the final conductance and can be the origin of the
opposite switching in the conductance mentioned in Fig. \ref{fig:TRC}.

\section{Conclusion} 

In this study we determined some possible paths of oxygen ions migration across
the MgO/V interface of the Fe/MgO/V/Fe junction. In particular, we have shown
that there is a local minimum of the total energy of the system as a function
of the oxygen position at the interface. This gives an interesting scenario
about how oxygen ions get trapped in a local minimum upon the application of a
strong electric field pulse and how the motion of oxygen get reversed when the
field is applied in the opposite direction and its connection to RSw. Moreover,
it is shown that switching is different depending on how the vanadium layer is
placed on top of the MgO layer (V on top of O or on top of Mg). This would
establish an easy experimental way of determining the structure of the
junction. In any case, the migration of the oxygen makes the RSw in MgO-based
MTJ more plausible. We hope that the mechanism presented in this work clarifies
somehow the RSw in MgO-based MTJ, and complements other model studies and will
be useful for the elaboration of new models based on realistic calculations.

We acknowledge support from an ANR pnano grant No. ANR-06-NANO-053-01. This
work was performed using HPC resources from GENSI-CINES Grant gem1100. We
acknowledge helpful discussions with I. Rungger and S. Sanvito.

% \newpage

% %%%%%% Bibliografy with bibtex
% 
% \bibliography{Refs/Refs.bib}

\begin{thebibliography}{10}
\bibitem{Sawa2008}
A. Sawa, \href{http://dx.doi.org/10.1016/S1369-7021(08)70119-6}{{\em Mater.
Today} {\bf 11}, 28 (2008).}

\bibitem{Waser2007}
R. Waser and M. Aono, \href{http://dx.doi.org/10.1038/nmat2023}{{\em Nature
Mater.} {\bf 6}, 833 (2007).}

\bibitem{Szot1992}
K. Szot, W. Speier and W. Eberhardt
\href{http://dx.doi.org/10.1063/1.107401}{{\em Appl. Phys. Lett.} {\bf 60},
1190 (1992).}

\bibitem{Yang2008}
J. J. Yang, M. D. Pickett, X. Li, D. A. A. Ohlberg, D. R. Stewart and R. S.
Williams, \href{http://dx.doi.org/10.1038/nnano.2008.160}{{\em Nature Nanot.}
{\bf 3}, 429 (2008).}

\bibitem{He2009}
C. L. He, F. Zhuge, X. F. Zhou, M. Li, G. C. Zhou, Y. W. Liu, J. Z. Wang, B.
Chen, W. J. Su, Z. P. Liu, Y. H. Wu, P. Cui and R.-W. Li,
\href{http://dx.doi.org/10.1063/1.3271177}{{\em Appl. Phys. Lett.} {\bf 95},
232101 (2009).}

\bibitem{Halley2008}
D. Halley, H. Majjad, M. Bowen, N. Najjari, Y. Henry, C. Ulhaq-Bouillet, W.
Weber, G. Bertoni, J. Verbeeck and G. Van Tendeloo,
\href{http://dx.doi.org/10.1063/1.2938696}{{\em Appl. Phys. Lett.} {\bf 92},
212115 (2008).}

\bibitem{Teixeira2009}
J. M. Teixeira, J. Ventura, R. Fermento, J. P. Araujo, J. B. Sousa, P.
Wisniowski and P. P. Freitas,
\href{http://dx.doi.org/10.1088/0022-3727/42/10/105407}{{\em J. Phys. D: Appl.
Phys.} {\bf 42}, 105407 (2009).}

\bibitem{Butler2001}
W. H. Butler, X.-G. Zhang, T. C. Schulthess and J. M. MacLaren,
\href{http://dx.doi.org/10.1103/PhysRevB.63.054416}{{\em Phys. Rev. B} {\bf
63}, 054416 (2001).}

\bibitem{Waser2009}
R. Waser, R. Dittmann, G. Staikov and K. Szot,
\href{http://dx.doi.org/10.1002/adma.200900375}{{\em Adv. Mater.} {\bf 21},
2632 (2009)}.

\bibitem{Simmons1967}
J. G. Simmons and R. R. Verderber,
\href{http://dx.doi.org/10.1098/rspa.1967.0191}{{\em Proc. R. Soc. Lond. A}
{\bf 301}, 77 (1967).}

\bibitem{Rozenberg2004}
M. J. Rozenberg, I. H. Inoue and M. J. Sanchez,
\href{http://dx.doi.org/10.1103/PhysRevLett.92.178302}{{\em Phys. Rev. Lett.}
{\bf 92}, 178302 (2004).}

\bibitem{Dearnaley1970}
G. Dearnaley, A. M. Stoneham and D. V. Morgan,
\href{http://dx.doi.org/10.1088/0034-4885/33/3/306}{{\em Rep. Prog. Phys.} {\bf
33}, 1129 (1970).}

\bibitem{Rozenberg2010}
M. J. Rozenberg, M. J. Sanchez, R. Weht, C. Acha, F. Gomez-Marlasca and P.
Levy,\href{http://dx.doi.org/10.1103/PhysRevB.81.115101}{{\em Phys. Rev. B}
{\bf 81}, 115101 (2010).}

\bibitem{Bertin2011}
E. Bertin, D. Halley, Y. Henry, N. Najjari, H. Majjad, M. Bowen, V. DaCosta, J.
Arabski and B. Doudin, \href{http://dx.doi.org/10.1063/1.3561497}{{\em J. Appl.
Phys.} {\bf 109}, 083712 (2011).}

\bibitem{Najjari2010}
N. Najjari, D. Halley, M. Bowen, H. Majjad, Y. Henry, and B. Doudin,
\href{http://dx.doi.org/10.1103/PhysRevB.81.174425}{{\em Phys. Rev. B} {\bf
81}, 174425 (2010).}

%===============================================
\bibitem{Bose2008}
 P. Bose,. A. Ernst, I. Mertig, and J. Henk, 
 \href{http://dx.doi.org/10.1103/PhysRevB.78.092403}{{\em Phys. Rev. B} {\bf
78}, 092403 (2008).}

\bibitem{Heiliger2007}
 C. Heiliger , P. Zahn, I. Mertig, 
 \href{http://dx.doi.org/10.1016/j.jmmm.2007.03.144}{{\em J. Magn. Magn. Mater.} {\bf 316}, 478 (2007).}

\bibitem{Wortmann2004}
 D. Wortmann, G. Bihlmayer and S. Blugel,
 \href{http://dx.doi.org/10.1088/0953-8984/16/48/056}{{\em J. Phys.: Condens. Matter} {\bf 16}, S5819 (2004).}

\bibitem{Heiliger2008}
 C. Heiliger, P. Zahn, B. Yu. Yavorsky, and I. Mertig,
 \href{http://dx.doi.org/10.1103/PhysRevB.77.224407}{{\em Phys. Rev. B} {\bf
77}, 224407 (2008).}

\bibitem{Ke2010}
 Youqi Ke, Ke Xia, and Hong Guo, \href{http://dx.doi.org/10.1103/PhysRevLett.105.236801}{{\em Phys. Rev. Lett.}
{\bf 105}, 236801 (2010).}
%===============================================

\bibitem{Chua1971}
L. O. Chua, \href{http://dx.doi.org/10.1109/TCT.1971.1083337}{{\em IEEE Trans.
Circuit Theory} {\bf 18}, 507 (1971).}

\bibitem{Feng2009}
X. Feng, O. Bengone, M. Alouani, I. Rungger and S. Sanvito,
\href{http://dx.doi.org/10.1103/PhysRevB.79.214432}{{\em Phys. Rev. B} {\bf
79}, 214432 (2009)}; X. Feng, O. Bengone, M. Alouani, S. Lebegue, I. Rungger
and S. Sanvito, \href{http://dx.doi.org/10.1103/PhysRevB.79.174414}{{\em Phys.
Rev. B} {\bf 79}, 174414 (2009).}

\bibitem{Ikuhara1997}
Y. Ikuhara, Y. Sugawara, I. Tanaka and P. Pirouz,
\href{http://dx.doi.org/10.1023/A:1008655609147}{{\em Interface Sci.} {\bf 5},
5 (1997).}

\bibitem{Rocha2006}
A. R. Rocha, V. M. Garcia-Suarez, S. Bailey, C. Lambert, J. Ferrer and S.
Sanvito, \href{http://dx.doi.org/10.1103/PhysRevB.73.085414}{{\em Phys. Rev. B}
{\bf 73}, 085414 (2006).}

\bibitem{Soler2002}
J. M. Soler, E. Artacho, J. D. Gale, A. Garcia, J. Junquera, P. Ordejon and D.
Sanchez-Portal, \href{http://dx.doi.org/10.1088/0953-8984/14/11/302}{{\em J.
Phys.: Condens. Matter} {\bf 14}, 2745 (2002).}

\bibitem{Troullier1991}
N. Troullier and J. L. Martins,
\href{http://dx.doi.org/10.1103/PhysRevB.43.1993}{{\em Phys. Rev. B} {\bf 43},
1993 (1991).}

\bibitem{Perdew1981}
J. P. Perdew and A. Zunger,
\href{http://dx.doi.org/10.1103/PhysRevB.23.5048}{{\em Phys. Rev. B} {\bf 23},
5048 (1981).}

\bibitem{Kurth2006}
M. Kurth, P. C. J. Graat and E. J. Mittemeijer,
\href{http://dx.doi.org/10.1016/j.tsf.2005.11.044}{{\em Thin Solid Films} {\bf
500}, 61 (2006).}

\end{thebibliography}
% %\bibliographystyle{jphysicsB} %epl style 
% %\bibliographystyle{unsrt} %epl style 
% \bibliographystyle{apsrev4-1} %revtex4
% %\bibliographystyle{plain}

\end{document}